\documentclass[12pt]{article}
\usepackage{amsmath}

\begin{document}

\begin{center}
\begin{large}
{\bf Composite system in noncommutative space and the equivalence principle}
\end{large}
\end{center}

\centerline { Kh. P. Gnatenko \footnote{E-Mail address: khrystyna.gnatenko@gmail.com }}
\medskip
\centerline {\small \it Ivan Franko National University of Lviv, Department for Theoretical Physics,}
\centerline {\small \it 12 Drahomanov St., Lviv, 79005, Ukraine}

\abstract{ The motion of a composite system made of N particles is examined in a space with a canonical noncommutative algebra of coordinates. It is found that the coordinates of the center-of-mass position satisfy noncommutative algebra with effective parameter. Therefore, the upper bound of the parameter of noncommutativity is re-examined. We conclude that the weak equivalence principle is violated in the case of a non-uniform gravitational field and propose the condition for the recovery of this principle in noncommutative space. Furthermore, the same condition is derived from the independence of kinetic energy on the composition.

 {\it Key words:} noncommutative space, composite system, equivalence principle.}

\section{Introduction}

Recently, noncommutativity has received much attention owing to the development of String Theory \cite{Connes,Witten} and Quantum Gravity \cite{Doplicher}. The idea that space might have a noncommutative structure has a long history. It was proposed by Heisenberg and was formalized by Snyder \cite{Snyder}.

The  noncommutative space can be realized as a space where the coordinate operators satisfy the following commutation relations
 \begin{eqnarray}
[\hat{X}_{i},\hat{X}_{j}]=i\hbar\theta_{ij},
\label{form1}
\end{eqnarray}
where $\theta_{ij}$ is a constant antisymmetric object. In the classical limit $\hbar\rightarrow0$ the quantum-mechanical commutator is replaced by the Poisson bracket
 \begin{eqnarray}
\{X_{i},X_{j}\}=\theta_{ij}.
\label{form2}
\end{eqnarray}

It is important to note that a charged and massive particle in a strong
magnetic field ${\bf B}$ pointing in the $Z$ direction moves in a noncommutative space. The commutation relation for the coordinates of a particle moving in the (X,Y) plane is given by
\begin{eqnarray}
[\hat{X},\hat{Y}]=-i\hbar\frac{c}{eB},
\end{eqnarray}
 here $e$ is the charge of the particle and $c$ is the speed of light \cite{Jackiw}.

Many physical problems have been studied in the framework of noncommutative quantum mechanics and noncommutative classical mechanics. Some of the first articles on quantum mechanics with noncommutativity of canonical type are \cite{Gamboa,Nair,Bolonek,Duval}. Formal aspects of noncommutative quantum mechanics are addressed in \cite{Bastos1,Bastos2}. Neutrons in a gravitational field with noncommutativity are considered in \cite{Bertolami}. Interesting effects arise when one considers noncommutativity in the context of quantum cosmology and black hole physics \cite{Bastos3,Garcia-Compean,Ansoldi}. The Landau problem \cite{Gamboa1,Horvathy,Dayi,Li,Dulat}, harmonic oscillator \cite{Hatzinikitas,Kijanka,Jing,Jahan}, two-dimensional system in central potential \cite{Gamboa}, classical particle in a gravitational potential \cite{Romero,Mirza}, classical systems with various potentials \cite{Djemai1} are studied. Note, however, that it is important to consider many-particle problem in order to analyze the properties of a wide class of physical systems in  noncommutative space.

The classical problem of many particles in noncommutative space-time was examined in \cite{ Daszkiewicz}. The authors considered two examples of many-particle systems, namely the set of N interacting harmonic oscillators and the system of N particles moving in the gravitational field. The corresponding Newton equation for each particle in these systems was provided.  In \cite{Jabbari} the two-body system of particles interacting through the harmonic oscillator potential was considered on a noncommutative plane. The authors implemented the noncommutativity through defining a new set of commutating coordinates and got the $\theta$-dependent Hamiltonian in usual commutative plane. The coordinates of the center-of-mass position and relative motion, the total momentum and the relative momentum were introduced in the traditional way. Therefore, the authors rewrote the Hamiltonian as a sum of the freely moving part and a $\theta$-dependent bounded term and derived the partition function of a two-body system of classical noncommutative harmonic oscillator.

The problems of noncommutative multiparticle quantum mechanics are examined in \cite{Ho}. The authors considered the case when the particles of opposite charges feel opposite noncommutativity. The coordinates of the center-of-mass and relative motion were introduced. It was shown that the magnitude of the center-of-mass coordinates noncommutativity is never large then the parameter of noncommutativity for elementary particle. In \cite{Djemai} a system of two quantum particles was considered in the context of noncommutative quantum mechanics, characterized by noncommutativity between the coordinates and momentum noncommutativity. The noncommutative correction to the energy spectrum of two-particle system was found. In \cite{Bellucci} the system of two charged quantum particles was considered in a space with coordinates noncommutativity. The authors reduced the two-body problem to a one-body problem for the internal motion.  The quantum model of many particles moving in twisted N-enlarged Newton-Hooke space-time was proposed in \cite{ Daszkiewicz1}. The Schroedinger equation for arbitrary stationary potential was provided. As an example the author examined the system of N particles moving "in" and interacting "by" the Coulomb potential.

 In the case of Doubly Special Relativity the problem of composite system (so-called soccer-ball problem) was considered in \cite{Girelli,Jacobson,Giovanni}. This problem was also studied within the framework of relative locality in \cite{Camelia,Hossenfelder,Camelia1}.

Composite system in deformed space with minimal length $[\hat{X},\hat{P}]=i\hbar(1+\beta\hat{P}^{2})$ was considered in \cite{Quesne}. The authors solved the two-body problem, studied the composite system made of N elementary particles, defined an effective deformation parameter and re-examined the estimation of the minimal length upper bound. In \cite{Tkachuk} the properties of the kinetic energy of a composite body were analyzed. The author considered the problem of violation of the equivalence principle and proposed the way to recover this principle in deformed space with minimal length.

 The violation of the equivalence principle is an important problem in noncommutative space. In \cite{Castello-Branco} the authors examined the free-fall of a quantum particle in a uniform gravitational field. It was argued that the equivalence principle extends to the realm of noncommutative quantum mechanics. One of the consequences of the twisted Poincare symmetry was investigated in \cite{Lee}. In this context the author concluded that one can expect that the equivalence principle is not violated in the noncommutative space-time.  However, in \cite{Bastos,Mehdipour,Saha}, the authors argued that noncommutativity leads to an apparent violation of the equivalence principle.

In this Letter the two-particle and N-particle systems are examined in noncommutative space. We consider the general case when the different particles satisfy noncommutative algebra with different parameters of noncommutativity. Every macroscopic body consist of elementary particles which feel the effect of noncommutativity with different parameters. So, there is a problem of describing the motion of the center-of-mass of macroscopic body in noncommutative space. In order to solve this problem the total momentum is introduced as an integral of motion in noncommutative space and the center-of-mass position is found as its conjugate variable. We conclude that the coordinates of the center-of-mass satisfy noncommutative algebra with effective parameter of noncommutativity. Taking into account this conclusion the condition to recover the equivalence principle in noncommutative space is proposed. Moreover, the same condition is derived from the independence of kinetic energy on the composition.

This Letter is organized as follows. In Section 2 the two-body problem is solved. More general case of composite system made of N elementary particles in noncommutative space is studied in Section 3. The motion of a body in gravitational field and the equivalence principle are considered in Section 4. The properties of the kinetic energy in noncommutative space are studied in Section 5. In Section 6 the upper bound of the parameter of  noncommutativity is re-examined.

\section{Two-body problem}

 In ordinary space we can reduce a two-body problem to an equivalent one-body problem. Let us consider two elementary particles of masses $m_{1}$, $m_{2}$  that interact only with each other in two-dimensional  noncommutative space and define the total momentum and the center-of-mass position of this system. We consider the case when the different particles of masses $m_{1}$, $m_{2}$  satisfy the noncommutative algebra with parameters $\theta_{1}$, $\theta_{2}$ respectively. Therefore, the coordinates $X_{\mu}^{(i)}$ and the components of momentum $P_{\mu}^{(i)}$ satisfy the following relations
\begin{eqnarray}
\{X_{1}^{(i)},X_{2}^{(j)}\}=-\{X_{2}^{(i)},X_{1}^{(j)}\}=\delta^{ij}\theta_{i},\\
\{X_{\mu}^{(i)},P_{\nu}^{(j)}\}=\delta_{\mu\nu}\delta^{ij},\\
\{P_{\mu}^{(i)},P_{\nu}^{(j)}\}=0,
\end{eqnarray}
here $\mu=(1,2)$, $\nu=(1,2)$ and the indices $i$, $j$ label the particles.

The interaction potential energy of the two particles  $V(|{\bf X}^{(1)}-{\bf X}^{(2)}|)$ depends on the distance between them. Therefore, the Hamiltonian of the system reads
\begin{eqnarray}
 H=\frac{({\bf P}^{(1)})^{2}}{2m_{1}}+\frac{({\bf P}^{(2)})^{2}}{2m_{2}}+V(|{\bf X}^{(1)}-{\bf X}^{(2)}|).
 \end{eqnarray}

  Let us introduce the total momentum in a traditional way
\begin{eqnarray}
\tilde{{\bf P}}={\bf P}^{(1)}+{\bf P}^{(2)}.
\label{form4}
\end{eqnarray}
It is easy to find that the total momentum satisfies the following relation
\begin{eqnarray}
\{\tilde{{\bf P}},H\}=0.
\end{eqnarray}

 So, the total momentum (\ref{form4}) is an integral of motion in noncommutative space. Now we can find the coordinates of the center-of-mass $\tilde{{\bf X}}$ as conjugate coordinates to the total momentum
\begin{eqnarray}
\tilde{{\bf X}}=\frac{m_{1}{\bf X}^{(1)}+m_{2}{\bf X}^{(2)}}{m_{1}+m_{2}}.
\end{eqnarray}
We can also introduce the coordinates and momentum of relative motion in the traditional way
\begin{eqnarray}
{\Delta\bf P}=\mu_{1}{\bf P}^{(2)}-\mu_{2}{\bf P}^{(1)},\label{form5}\\
{\Delta\bf X}={\bf X}^{(2)}-{\bf X}^{(1)},
\end{eqnarray}
 here $\mu_{1}=m_{1}/(m_{1}+m_{2})$ and $\mu_{2}=m_{2}/(m_{1}+m_{2})$.

It is easy to find that
  \begin{eqnarray}
\{\tilde{X}_{\mu},\tilde{P}_{\nu}\}=\{\mu_{1}X_{\mu}^{(1)}+\mu_{2}X_{\mu}^{(2)},P_{\nu}^{(1)}+P_{\nu}^{(2)}\}=\delta_{\mu\nu},\\
 \{\Delta X_{\mu},\Delta P_{\nu}\}=\{X_{\mu}^{(2)}-X_{\mu}^{(1)},\mu_{1}P_{\nu}^{(2)}-\mu_{2}P_{\nu}^{(1)}\}=\delta_{\mu\nu},\\
 \{\tilde{P}_{\mu},\tilde{P}_{\nu}\}=\{\Delta P_{\mu},\Delta P_{\nu}\}=0.
 \end{eqnarray}

Let us calculate the Poisson brackets for the coordinates of the center-of-mass
 \begin{eqnarray}
\{\tilde{X}_{1},\tilde{X}_{2}\}=-\{\tilde{X}_{2},\tilde{X}_{1}\}=\frac{1}{(m_{1}+m_{2})^{2}}\{m_{1}X_{1}^{(1)}+m_{2}X_{1}^{(2)}, m_{1}X_{2}^{(1)}+m_{2}X_{2}^{(2)}\}=\frac{m_{1}^{2}\theta_{1}+m_{2}^{2}\theta_{2}}{(m_{1}+m_{2})^{2}}.
\end{eqnarray}

So, to describe the motion of the center-of-mass we need to introduce an effective parameter of noncommutativity
\begin{eqnarray}
\tilde{\theta}=\frac{m_{1}^{2}\theta_{1}+m_{2}^{2}\theta_{2}}{(m_{1}+m_{2})^{2}}.
\label{form7}
\end{eqnarray}

The relative coordinates also satisfy noncommutative algebra relations with an effective parameter $\theta_{1}+\theta_{2}$
\begin{eqnarray}
\{\Delta X_{1},\Delta X_{2}\}=-\{\Delta X_{2},\Delta X_{1}\}=\{X_{1}^{(2)}-X_{1}^{(1)},X_{2}^{(2)}-X_{2}^{(1)}\}=\theta_{1}+\theta_{2}.
\label{form8}
\end{eqnarray}

Note, that the coordinates of the center-of-mass position and the coordinates of relative motion are not independent. It is easy to find that
\begin{eqnarray}
\{\tilde{X}_{1},\Delta X_{2}\}=\left\{\frac{m_{1}X_{1}^{(1)}+m_{2}X_{1}^{(2)}}{m_{1}+m_{2}},X_{2}^{(2)}-X_{2}^{(1)}\right\}=\frac{m_{2}\theta_{2}-m_{1}\theta_{1}}{m_{2}+m_{1}},\label{form9}\\
\{\Delta X_{1},\tilde{X}_{2}\}=\left\{X_{1}^{(2)}-X_{1}^{(1)},\frac{m_{1}X_{2}^{(1)}+m_{2}X_{2}^{(2)}}{m_{1}+m_{2}}\right\}=\frac{m_{2}\theta_{2}-m_{1}\theta_{1}}{m_{2}+m_{1}}.
\label{form10}
\end{eqnarray}

So, there is an influence between the motion of the center-of-mass and the relative motion in noncommutative space. Note, that we can avoid this noncommutativity (\ref{form9}), (\ref{form10}) in the cases of $m_{1}=m_{2}$, $\theta_{1}=\theta_{2}$ and $m_{i}\theta_{i}=\gamma=const$, $i=(1,2)$.

The Hamiltonian of two-particle system becomes
\begin{eqnarray}
H=\frac{(\tilde{{\bf P}})^{2}}{2M}+\frac{(\Delta{\bf P})^{2}}{2\mu}+V(|\Delta{\bf X}|),
\end{eqnarray}
here $M=m_{1}+m_{2}$ is the total mass and $\mu=m_{1}m_{2}/(m_{1}+m_{2})$ is the reduced mass.
So, the two-particle problem can be reduced to a one-particle problem for the internal motion in  noncommutative space.

It is important to note that it is easy to generalize this classical problem to the quantum case.

 \section{Composite system made of N elementary particles in noncommutative space}

Let us consider a more general case. We have a system made of N elementary particles of masses $m_{i}$ that interact with each other in two-dimensional noncommutative space. The N particles Hamiltonian reads
 \begin{eqnarray}
  H=\sum_{i}\frac{({\bf P}^{(i)})^{2}}{2m_{i}}+\frac{1}{2}\mathop{\sum_{i,j}}\limits_{i\neq j} V(|{\bf X}^{(i)}-{\bf X}^{(j)}|),
 \end{eqnarray}
 here $i$ and $j$ run over $1,2,...,N$.

 The coordinates $X_{\mu}^{(i)}$ and the components of momentum $P_{\mu}^{(i)}$ satisfy the following noncommutative algebra relations
  \begin{eqnarray}
 \{X_{1}^{(i)},X_{2}^{(j)}\}=-\{X_{2}^{(i)},X_{1}^{(j)}\}=\delta^{ij}\theta_{i},\\
\{X_{\mu}^{(i)},P_{\nu}^{(j)}\}=\delta_{\mu\nu}\delta^{ij},\\
\{P_{\mu}^{(i)},P_{\nu}^{(j)}\}=0,
 \end{eqnarray}
  here $\theta_{i}$ is the parameter of noncommutativity that corresponds to the particle with mass $m_{i}$ $(i=1,2,...,N)$.

Let us introduce the total momentum in the traditional way $\tilde{{\bf P}}=\sum_{i}{\bf P}^{(i)}.$  It is easy to show that $\{\tilde{{\bf P}},H\}=0$. Therefore, the $\tilde{{\bf P}}$ is an integral of motion in  noncommutative space. So, we can define the coordinates of the center-of-mass position, the momentum and the coordinates of relative motion in the traditional way
\begin{eqnarray}
\tilde{{\bf X}}=\sum_{i}\mu_{i}{\bf X}^{(i)},\\
\Delta{\bf P}^{(i)}={\bf P}^{(i)}-\mu_{i}\tilde{{\bf P}},\\
\Delta{\bf X}^{(i)}={\bf X}^{(i)}-\tilde{{\bf X}},
\end{eqnarray}
here $\mu_{i}=m_{i}/M$, $M=\sum_{i}m_{i}$. It is easy to find that
 \begin{eqnarray}
\{\tilde{X}_{\mu},\tilde{P}_{\nu}\}=\delta_{\mu\nu},\\
\{\Delta X^{(i)}_{\mu},\Delta P^{(j)}_{\nu}\}=\delta_{\mu\nu}(\delta^{ij}-\mu_{j}),\\
\{\tilde{P}_{\mu},\tilde{P}_{\nu}\}=\{P^{(i)}_{\mu},P^{(j)}_{\nu}\}=0.
 \end{eqnarray}

Let us calculate the Poisson brackets for the coordinates of the center-of-mass
\begin{eqnarray}
\{\tilde{X}_{1}, \tilde{X}_{2}\}=-\{\tilde{X}_{2}, \tilde{X}_{1}\}=\sum_{i}\sum_{j}\{\mu_{i}X_{1}^{(i)},\mu_{j}X_{2}^{(j)}\}=\frac{\sum_{i}m_{i}^{2}\theta_{i}}{(\sum_{l}m_{l})^{2}}.
\end{eqnarray}

So, the coordinates of the center-of-mass satisfy the noncommutative algebra with effective parameter of noncommutativity
\begin{eqnarray}
\tilde{\theta}=\frac{\sum_{i}m_{i}^{2}\theta_{i}}{(\sum_{j}m_{j})^{2}}.
\label{form11}
\end{eqnarray}
Therefore, the parameter of noncommutativity for a macroscopic body depends on its composition. Note, however, that in the case of $m_{i}\theta_{i}=\gamma$ we find
\begin{eqnarray}
 \tilde{\theta}=\frac{\gamma}{M},
 \label{form12}
 \end{eqnarray}
 and thus the effective parameter of noncommutativity does not depend on the composition.

In the case of $m_{1}=m_{2}=...=m_{N}$ and $\theta=\theta_{1}=\theta_{2}=...=\theta_{N}$ the effective parameter of noncommutativity is given by
\begin{eqnarray}
\tilde{\theta}=\frac{\theta}{N}.
\end{eqnarray}

So, the value of reduction of the effective parameter with respect to the parameter of noncommutativity for the individual particles depends on the number of particles in the system (macroscopic body).

Note, that the coordinates of relative motion satisfy nontrivial relations
 \begin{eqnarray}
 \{\Delta X^{(i)}_{1}, \Delta X^{(j)}_{2}\}=-\{\Delta X^{(i)}_{2}, \Delta X^{(j)}_{1}\}=\delta^{ij}\theta_{i}-\mu_{i}\theta_{i}-\mu_{j}\theta_{j}+\tilde{\theta}.
  \end{eqnarray}

The coordinates of the center-of-mass and the relative coordinates are not independent in noncommutative space. It is easy to find that
\begin{eqnarray}
\{\tilde{X}_{1}, \Delta X_{2}^{(i)}\}=\mu_{i}\theta_{i}-\tilde{\theta},\\
\{\Delta X_{1}^{(i)}, \tilde{X}_{2}\}=\mu_{i}\theta_{i}-\tilde{\theta}.
\end{eqnarray}

Note, however, that the coordinates become independent in the case of $m_{i}\theta_{i}=\gamma=const$.

In the next section we will show that  expression (\ref{form11}) is helpful to find the way to recover the equivalence principle in  noncommutative space.

\section{The motion of a body in a gravitational field. The equivalence principle}

 In this section we examine the motion of a macroscopic body which we consider as a point particle in uniform and non-uniform gravitational fields.

 Let us first consider free fall of a macroscopic body of mass $M$ with effective parameter of noncommutativity $\tilde{\theta}$ in a uniform gravitational field. In the case when the free fall acceleration ${\bf g}$ is directed along the $X$ axis the Hamiltonian of the body reads
\begin{eqnarray}
H=\frac{P_{x}^{2}}{2M}+\frac{P_{y}^{2}}{2M}-MgX.
\end{eqnarray}
Taking into account the  noncommutative algebra relations
\begin{eqnarray}
\{X,Y\}=\tilde{\theta},\\
\{X,P_{x}\}=1,\\
\{Y,P_{y}\}=1,\\
\{P_{x},P_{y}\}=0,
\end{eqnarray}
we obtain the following equations of motion
\begin{eqnarray}
\dot{X}=\{X,H\}=\frac{P_{x}}{M},\label{form13}\\
\dot{Y}=\{Y,H\}=\frac{P_{y}}{M}+Mg\tilde{\theta},\label{form14}\\
\dot{P_{x}}=\{P_{x},H\}=Mg,\label{form15}\\
\dot{P_{y}}=\{P_{y},H\}=0.
\label{form16}
\end{eqnarray}

Let us introduce initial conditions
\begin{eqnarray}
\left\{
\begin{aligned}
X(0)=X_{0},\\
Y(0)=Y_{0},\\
\dot{X}(0)=\upsilon_{ox},\\
\dot{Y}(0)=\upsilon_{oy}.
\end{aligned}
\right.
\label{form17}
\end{eqnarray}
The solutions of equations (\ref{form13})-(\ref{form16}) with initial conditions (\ref{form17})
read
\begin{eqnarray}
X(t)=\frac{gt^{2}}{2}+\upsilon_{ox}t+X_{0},\label{form18}\\
Y(t)=\upsilon_{oy}t+Y_{0},\label{form19}\\
P_{x}(t)=Mgt+M\upsilon_{ox},\\
P_{y}(t)=-M^{2}g\tilde{\theta}+M\upsilon_{oy}.
\end{eqnarray}

So, we obtain the trajectory of a body in noncommutative space. The weak equivalence principle (also called the uniqueness of free fall principle or the Galilean equivalence principle) states the trajectory of a point mass in a gravitational field depends only on its initial position and velocity, and is independent of its composition and structure. This principle is a restatement of the equality of gravitational and inertial mass. Therefore, taking into account equations (\ref{form18}), (\ref{form19}), we conclude that the weak equivalence principle is not violated in this case.

Let us generalize this problem into the case of non-uniform gravitational field $V(X,Y)$. The Hamiltonian of a body of mass $M$ in this field reads
\begin{eqnarray}
 H=\frac{P_{x}^{2}}{2M}+\frac{P_{y}^{2}}{2M}+MV(X,Y).
 \end{eqnarray}

The equations of motion in  noncommutative space can be written as follows
 \begin{eqnarray}
 \dot{X}=\{X,H\}=\{X,\frac{P_{x}^{2}}{2M}+\frac{P_{y}^{2}}{2M}+MV(X,Y)\}=\frac{P_{x}}{M}+M\tilde{\theta}\frac{\partial V(X,Y)}{\partial Y},\label{form20}\\
\dot{Y}=\{Y,H\}=\{Y,\frac{P_{x}^{2}}{2M}+\frac{P_{y}^{2}}{2M}+MV(X,Y)\}=\frac{P_{y}}{M}-M\tilde{\theta}\frac{\partial V(X,Y)}{\partial X},\label{form21}\\
\dot{P_{x}}=\{P_{x},H\}=\{P_{x},\frac{P_{x}^{2}}{2M}+\frac{P_{y}^{2}}{2M}+MV(X,Y)\}=-M\frac{\partial V(X,Y)}{\partial X},\\
\dot{P_{y}}=\{P_{y},H\}=\{P_{y},\frac{P_{x}^{2}}{2M}+\frac{P_{y}^{2}}{2M}+MV(X,Y)\}=-M\frac{\partial V(X,Y)}{\partial Y}.
\end{eqnarray}

Note, that the equations of motion (\ref{form20}), (\ref{form21}) depend on the mass of the body and its effective parameter of noncommutativity. So, the weak equivalence principle is violated in this case.

It is important to note that according to (\ref{form11}) the effective parameter of noncommutativity depends on the composition of a body. Therefore, in the case of different bodies of the same masses but with different compositions the equivalence principle is also violated. As was shown in the Section 3 the effective parameter of noncommutativity does not depends on the composition of a body in the case of $m_{i}\theta_{i}=\gamma=const$. Substituting (\ref{form12}) into equations (\ref{form20}) and (\ref{form21}) we obtain
  \begin{eqnarray}
\left\{
\begin{aligned}
 \dot{X}=\frac{P_{x}}{M}+\gamma\frac{\partial V(X,Y)}{\partial Y},\\
\dot{Y}=\frac{P_{y}}{M}-\gamma\frac{\partial V(X,Y)}{\partial X},\\
\dot{P_{x}}=-M\frac{\partial V(X,Y)}{\partial X},\\
\dot{P_{y}}=-M\frac{\partial V(X,Y)}{\partial Y}.
\end{aligned}
\right.
\label{form22}
\end{eqnarray}

The equations of motion (\ref{form22}) depend on the $\gamma$ constant which is the same for all bodies.
So, we can recover the equivalence principle in the case of
\begin{eqnarray}
m_{i}\theta_{i}=\gamma=const.
\label{form23}
\end{eqnarray}

Note, that the mass and the effective parameter of the macroscopic body also satisfy this condition
\begin{eqnarray}
M\tilde{\theta}=\gamma=const.
\end{eqnarray}
Moreover, expression (\ref{form23}) can be derived from the independence of kinetic energy on the composition of a body.

\section{Kinetic energy in noncommutative space}

The kinetic energy of a body of mass $M$ reads
\begin{eqnarray}
T =\frac{P_{x}^{2}}{2M}+\frac{P_{y}^{2}}{2M}.
\end{eqnarray}
Let us consider the additivity property of the kinetic energy and its independence of
the composition of a body.

Taking into account (\ref{form13}), (\ref{form14}), we can rewrite the kinetic energy as a function of velocity in the first-order approximation over $\tilde{\theta}$.
\begin{eqnarray}
T=\frac{M\dot{X}^{2}}{2}+\frac{M\dot{Y}^{2}}{2}-M^{2}\tilde{\theta}g\dot{Y}.
\label{form24}
\end{eqnarray}

Let us assume that the body can be divided into N parts of masses $m_{i}$ and parameters of noncommutativity $\theta_{i}$  which can be treated as point particles. The kinetic energy of a particle reads

\begin{eqnarray}
T_{i}=\frac{(P_{x}^{(i)})^{2}}{2m_{i}}+\frac{(P_{y}^{(i)})^{2}}{2m_{i}}=\frac{m_{i}(\dot{X}^{(i)})^{2}}{2}+\frac{m_{i}(\dot{Y}^{(i)})^{2}}{2}-m_{i}^{2}\theta_{i}g\dot{Y}^{(i)}.
\end{eqnarray}

 Let us consider the case when the velocity of each particle in the system is the same as the velocity of the whole system $\dot{X}^{(i)}=\dot{X}$, $\dot{Y}^{(i)}=\dot{Y}$. According to the additivity property of the kinetic energy, we obtain
\begin{eqnarray}
T=\sum_{i}T_{i}=\frac{M\dot{X}^{2}}{2}+\frac{M\dot{Y}^{2}}{2}-\sum_{i}m_{i}^{2}\theta_{i}g\dot{Y}.
\label{form25}
\end{eqnarray}
 Comparing (\ref{form24}) and (\ref{form25}), we find (\ref{form11}). Note that the kinetic energy of a body (\ref{form24}) depends on the effective parameter of noncommutativity $\tilde{\theta}$ and therefore depends on its composition.

 Let us consider a body of a fixed mass $M$ consisting of two parts of masses $m_{1}$ and $m_{2}$.  Taking into account (\ref{form7}), we can write the effective parameter of noncommutativity in the following form
\begin{eqnarray}
\tilde{\theta}=\frac{m_{1}^{2}\theta_{1}+m_{2}^{2}\theta_{2}}{(m_{1}+m_{2})^{2}}=\theta_{\mu}\mu^{2}+\theta_{1-\mu}(1-\mu)^{2},
\label{form26}
\end{eqnarray}
  here $\theta_{\mu}=\theta_{1}$, $\theta_{1-\mu}=\theta_{2}$, $\mu=m_{1}/M$ and $1-\mu=m_{2}/M$.

 According to (\ref{form24}) the kinetic energy does not depend on the composition in the case of  $\tilde{\theta}=const$  for different bodies of equal mass but with different composition. Therefore, we can find $\theta_{\mu}$ as a function of $\mu$ from equation (\ref{form26})
\begin{eqnarray}
\theta_{\mu}=\frac{\tilde{\theta}}{\mu}.
\label{form27}
\end{eqnarray}
Taking into account that $\mu=m_{1}/M$ and $1-\mu=m_{2}/M$, we obtain $m_{1}\theta_{1}=m_{2}\theta_{2}=M\tilde{\theta}=const$.

So, we have derived expression (\ref{form23}) from the condition of
independence of kinetic energy on the composition.

 \section{The upper bound of the parameter of noncommutativity}

The motion of a particle in a gravitational potential was considered in \cite{Romero,Mirza,Djemai1}. The authors defined the perihelion shift caused by the coordinates noncommutativity. In the particular case of the Mercury planet by comparing the perihelion shift to the experimental data, the upper bound of the parameter of noncommutativity was found.

It is important to note that the authors in  \cite{Romero,Mirza,Djemai1} did not take into account that the macroscopic body, such as Mercury, feels coordinates noncommutativity with the effective parameter. So, the upper bounds of the parameters of noncommutativity obtained in these articles have to be corrected.
 For deformed nonlinear algebra with minimal length such correction was done in \cite{Quesne}. In this Letter we find the upper bound of the parameter of noncommutativity for nucleons considering the motion of Mercury in noncommutative space.

In order to re-examine these upper bounds estimation let us find the effective parameter of noncommutativity for the Mercury. The main contribution to the mass of the planet comes from the nucleons (neutrons and protons). So, we can calculate the number of nucleons with the help of the following relation
  \begin{eqnarray}
 N_{nuc}\simeq\frac{M}{m_{nuc}}=1.98\cdot10^{50},
 \end{eqnarray}
 here $M=3.3\cdot10^{23}kg$ is the mass of Mercury, $m_{nuc}=1.67\cdot10^{-27}kg$ is the mass of nucleons. The number of electrons is the same as the number of protons and approximately equal $\frac{1}{2}N_{nuc}$.
 Note that
 \begin{eqnarray}
 \frac{m_{e}}{M}\simeq\frac{m_{e}}{N_{nuc}m_{nuc}}\simeq\frac{1}{1840N_{nuc}}.
 \label{form28}
 \end{eqnarray}
Taking into account (\ref{form11}) we find
   \begin{eqnarray}
   \tilde{\theta}=N_{nuc}\theta_{nuc}(\frac{m_{nuc}}{M})^{2}+N_{e}\theta_{e}(\frac{m_{e}}{M})^{2}.
   \label{form29}
    \end{eqnarray}
    The nucleons are made of three quarks. Therefore, $\tilde{\theta}_{nuc}=\theta_{q}/3$,
 here $\theta_{q}$ is the parameter of noncommutativity for quarks.
Assuming that $\theta_{q}=\theta_{e}$ and taking into account (\ref{form28}), the
second term in (\ref{form29}) may be omitted. So, the effective parameter can be written as follows  \begin{eqnarray}
\tilde{\theta}=\frac{\theta_{nuc}}{N_{nuc}}.
 \label{form30}
\end{eqnarray}

 Taking into account (\ref{form30}), we can re-examine the upper bounds obtained in \cite{Romero,Mirza,Djemai1}. The authors found  $\hbar\theta\leq21\cdot10^{-64}m^{2}$ \cite{Romero}, $\hbar\theta\leq10^{-62}m^{2}$ \cite{Mirza}, and $\hbar\theta\leq40\cdot10^{-62}m^{2}$ \cite{Djemai1}. Therefore we obtain  $\hbar\theta_{nuc}\leq4.2\cdot10^{-13}m^{2}$, $\hbar\theta_{nuc}\leq2\cdot10^{-12}m^{2}$ and  $\hbar\theta_{nuc}\leq7.9\cdot10^{-11}m^{2}$ respectively.

These results are closer to the result obtained in \cite{Castello-Branco}. By resorting to experimental data from the GRANIT experiment, in which the first energy levels of freely falling neutrons were determined, the authors impose an upper bound of the parameter of noncommutativity $\hbar\theta\leq 0.771\cdot10^{-13}m^{2},(n = 1)$, $\hbar\theta\leq 1.021\cdot10^{-13}m^{2},(n = 2)$.

\section{Conclusion}

In this Letter we have examined the motion of a composite system made of N particles in a noncommutative space. The two-body problem has been considered. We have introduced the coordinates of the center-of-mass, the total momentum, the coordinates and the momentum of relative motion.  Therefore, the two-body problem has been reduced to an equivalent one-body problem.

 An effective parameter of noncommutativity has been introduced to describe motion of a composite system
\begin{eqnarray}
\tilde{\theta}=\frac{\sum_{i}m_{i}^{2}\theta_{i}}{(\sum_{j}m_{j})^{2}},
\end{eqnarray}
where $i$ and $j$ run over $1,2,...,N$. It has been shown that in the case of $m_{1}=m_{2}=...=m_{N}$ and $\theta_{1}=\theta_{2}=...=\theta_{N}=\theta$ the effective parameter of noncommutativity is given by $\tilde{\theta}=\theta/N$. So, the value of reduction of the effective parameter with respect to the parameter of noncommutativity for the individual particles depends on the number of particles in the system. Therefore, the upper bounds of the parameter of noncommutativity obtained for the macroscopic body (the Mercury planet) have been re-examined.

We have concluded that the coordinates of the center-of-mass and relative motion do not commute in  noncommutative space. So, they are not independent. Note, however, that we can avoid an influence between the motion of the center-of-mass and the relative motion in the case of
 \begin{eqnarray}
     \theta_{i}=\frac{\gamma}{m_{i}},
       \label{form31}
       \end{eqnarray}
where $\gamma=const$.

It has been shown that the same condition (\ref{form31}) gives the possibility to solve the problem of violation of the equivalence principle in  noncommutative space. Namely we have shown that the equivalence principle is recovered on condition (\ref{form31}).

Furthermore, expression (\ref{form31}) has been derived from the condition of the independence of kinetic energy on the composition.

 So, at least three problems caused by the coordinates noncommutativity can be solved by assuming that the parameter of noncommutativity is determined by the mass of a particle (\ref{form31}).

It is important to note that the $\gamma$ constant has a time dimension. In order to estimate the value of it let us suppose that the parameter of  noncommutativity for the electron $\theta_{e}$ satisfies the following relation
\begin{eqnarray}
\sqrt{\hbar\theta_{e}}=L_{p},
 \label{form32}
\end{eqnarray}
 where $L_{p}$ is the Planck length. Therefore, we find
 \begin{eqnarray}
 \gamma=m_{e}\theta_{e}=\frac{m_{e} L_{p}^{2}}{\hbar}=2\pi\frac{L_{p}}{\lambda_{e}}T_{p}=4.2\cdot10^{-23}T_{p}=2.3\cdot10^{-66}s,
 \end{eqnarray}
here $\lambda_{e}=h/m_{e}c=2.43\cdot10^{-12}m$ is Compton wavelength of the electron and $T_{p}=5.4\cdot10^{-44}s$ is the Planck time.

 In this case the parameter of noncommutativity $\theta_{i}$ for a particle of mass $m_{i}$ reads
 \begin{eqnarray}
 \hbar\theta_{i}=\hbar\theta_{e}\frac{m_{e}}{m_{i}}=L_{p}^{2}\frac{m_{e}}{m_{i}}.
 \end{eqnarray}

 So, the value of reduction of the parameter of noncommutativity for a particle of mass $m_{i}$  with respect to the parameter of noncommutativity for the electrons depends on the ratio $m_{e}/m_{i}$. Note, that in the case of $m_{i}=m_{nuc}$ we find $\hbar\theta_{nuc}=L_{p}^{2}/1840$.

\medskip
\begin{large}
{\bf Acknowledgments}
\end{large}

 The author thanks Professor V. M. Tkachuk for his advices and great support during research studies.

\end{document}